\documentstyle[proceedings,numreferences]{crckapb}
\begin{opening}
\title{STRATEGIES FOR THE SEARCH OF LIFE IN THE UNIVERSE}
\author{J. SCHNEIDER}
\institute{CNRS\\
Observatoire de Paris, 92195 Meudon, France}
\end{opening}
\begin{document}

\section{Introduction}
The question of the existence of Life elsewhere in the Universe is 
probably the oldest scientific question since it was already formulated in 
appropriate terms by Epicurus in his  `Letter to Herodotus' \cite{Epicurus}.
Today, particularly with the opportunities opened by space missions, this 
perspective is becoming a scientific reasonable goal. Nevertheless one 
prerequisit is to have an idea of what we want to search for, i.e. an idea
of what we mean by `Life'. In order to miss the fewest conceivable
forms of Life, we should start with as few {\it a priori} assumptions as 
possible. Let us briefly sketch an argumentation in 6 steps:
\begin{enumerate}
\item  Contrary to what is usually claimed by biologists, the very essence of 
Life is a subjective notion. `Life' is, at first, only recognized as such on 
{\it a priori subjective} grounds. The everyone's experience shows the 
importance of a priviledged criterion: richness of subjective exchanges 
with {\it us}.

\item It is only in a second step that the physical analysis of a system
judged as living exhibits an essential attribute: it has
rich physical exchanges with its surrounding, implying that its structure is 
sufficiently complex.
Physically speaking, a complex system is an out of thermodynamical 
equilibrium system.

\item Out of equilibrium systems can spontaneously acquire a complex structure 
only after a long natural process of self-complexification (amplification
of instabilities). 
In biological terms, this means {\it Evolution}.
This requires a {\it permanent source of high neguentropy}.

\item Because of the second law of thermodynamics, a complex system is unstable.
Thus there must be a continuous {\it regeneration} mechanism which 
maintains the complex structure. This mechanism necessarily requires 
{\it exchanges} of matter and energy with the {\it surrounding}. 
In biological terms, this means alimentation and photosynthesis

\item If alimentation is not sufficient, there is another mechanism named
{\it reproduction}. On the Earth, we know two such reproduction mechanisms: 
mitosis and sexual reproduction. 

\item Another reasonnable requirement is that, in order to minimize the 
expenses in energy, the structure of the system should be as stable as 
possible, or its evolution as steady as possible. 
This is well implemented by 
some kind of {\it memory}, together with a reading mechanism. 
The  genetic code is a good example of such a memory.
On the other hand, the stability of the systems also requires the 
{\it stability of the environment}.
\end{enumerate}
The above requirements are too general to serve as guidelines for actual
searches of Life outside the solar system. We need more specific hypotheses.
\begin{enumerate}
\item What kind of complex system? There are several possibilities:
      \begin{enumerate}
      \item chemical (organic) systems
      \item electromagnetic plasmas
      \item solid state physics
      \item liquid electronics, liquid crystals ....
      \item others ? ...
      \end{enumerate}
A standard conservative choice is to restrict the search to chemical systems.
We can then specify further more restricting hypotheses:
\item Consider only {\it carbon-based organic chemistry}
\item Require the presence of {\it liquid(s)}: favors the convective and 
hydrodynamical mixing of molecules
\item The liquid must be  {\it water}: it is a very good dissolvent and
      is abundant in nature.
\item Require the existence of a {\it solid/liquid interface} to enhance the
exchanges between molecules.
\end{enumerate}
From the above `decision tree', the safest environment which our imagination 
has yet found is an `habitable' planet. It is a planet having the following 
characterisctics:\\
- It is in orbit around a `main sequence' star (i.e. a star burning steadily 
its light elements into heavy elements) for its source of 
out-of-equilibrium photons. \\
- It must be a solid planet to allow for a liquid-solid interface; this
excludes (giant) gaseous planets of the Jupiter type.\\
- It is at the right distance from the star to allow for liquid water.
A planet orbiting at a distance $a$ around a star with a radius $R_*$ and a 
temperature $T_*$ acquires, by heating an equilibrium temperature $T_P$ given by
\begin{equation}
T_P = \frac{(1-A)^{1/4}}{\sqrt2}\left( \frac{R_*}{a}\right) ^{1/2}T_*
\end{equation}
where $A$ is the mean `albedo' (reflectance) of the planet surface.
For instance, by Eq. (1), the equilibrium temperature of the Earth ($A=.39$),
heated by the Sun ($T_{\odot} = 5770 K$), is
\begin{displaymath}
T_{\oplus} =  280 K
\end{displaymath}
(the actual mean temperature of the Earth's atmosphere is 287 K).\\

Thus, from Eq. (1), a planet having a temperature $\simeq 300 \pm 20 K$ (to 
allow for liquid water) must be located at a distance $a$
from the star given by (for $A=1$):
\begin{equation}
a=R_*\left(\frac{T_*}{300}\right) ^2
\end{equation}
Depending on the type of the central star, this distance runs from 
$\simeq$ 0.1 AU for cool ($T_*=3000K$) dwarf stars to $\simeq$ 2 AU for
hot ($T_* = 6500K$) larger stars (1 AU, or Astronomical Unit is the Sun-Earth 
distance). Hoter stars evolve too rapidly to have stable temperature conditions.

\section{Detection of habitable planets: what it is conceivable to do.}
The potential success of a given detection method 
depends naturally on its technological limitations, but also on the different
characteristics of the planets: their mass $M$, radius $R$
and distance $a$ from the parent star (I assume circular 
orbits). 

\subsection{The observables in each detection method.}
Consider a planetary system at a distance $D$ from the solar system with a
planet at a distance $a$ from its `parent' star. There are several methods to 
detect the planet either directly or indirectly. 
\begin{enumerate}
\item The most starightforward idea is to take a picture of the planet.
Unfortunately it is at a very small angular distance $\Delta \theta =a/D$ 
from the star
and is, by illumination by the star, very faint (the planet to
star brightness ratio $I_P/I_* =(1/4) (R_P/a)^2$ is only 
$\simeq 10^{-9}-10^{-10}$). Its image is therefore blurried by the star's halo
due to the diffraction and diffusion of light in the telescope.
\item A second series of methods rests on the detection of the `reflex' motion 
of the star due to the gravitational influence of the planet: the star goes,
with a period equal to the period $P$ of orbital revolution of the planet, over 
an orbit with a radius $(M_P/M_*)\times a$ around the center of mass of the 
planet-star system. The star's motion can be detected by three means:
  \begin{enumerate}
  \item Modulation of the star's position on the sky: its apparent trajectory
        is the elliptic projection of its orbit and has an angular
        amplitude $\Delta \alpha =(M_P/M_*)(a/D)$.
  \item Measurement of the star's velocity: by application of the Kepler laws,
        the velocity of the star is modulated with a period $P$ and an
        amplitude $V_R=(M_P/M_*)\sqrt{GM_*/a}$, where $G$ is the Newton 
        constant of gravity.
  \item Variation of the star's distance to the observer. Suppose there is
        some emission mechanism on the star, sending periodic signals to the 
        observer. When the star moves, the observer will receive the signals 
        with a (positive or negative) delay with respect to their mean time of
        arrival because of the variation of length of the star-observer 
        distance. This delay will be modulated with a period $P$ and an
        amplitude $\Delta T = (M_P/M_*)(a/c)$, where $c$ is the speed of light.
        In actual situations, the star can be a pulsar and the periodic signal
        the pulse of the pulsar.
  \end{enumerate}
\item If the star has modulations in 
its light curve (flares etc.), the planet presents, by illumination,
the same modulations with a delay $DT$ and a relative amplitude $R^2/a^2$.
An autocorrelation study of the star light curve could reveal this echo
\cite{Argy}. 
\item If the orbital plane of the planet is correctly oriented, it
produces a drop $\eta$ in the star light during transits 
(duration $D_{occ}\approx$ 1-20hrs);
for a random orientation of the planet orbital plane, the geometric probability
of the transit is $p=R_{\star}/a$ for a single star; for an
eclipsing binary this probability can amount 100\% since it is likely that
the binary and planet orbital planes are identical; 
in the latter case the transit is double (\cite{Sch1}).
\item The planet can produce a gravitationnal amplification $A_G$ of the light of 
background stars with a duration $T_G$ depending on its transverse velocity $V$
\cite{Pac}. 
\item Finally the planet can, like Jupiter or Saturn, have 
an intrinsic radio emission with a flux $F_R$ and could be searched by this
emission \cite{Lec1}.
At the present knowledge, the latter is
possible for any kind of planet and thus is not necessarily related to other 
planet characteristics such as $a$ or $M$; for instance, in the solar system,
the Earth is brighter than Uranus and Neptune at low radio frequencies
\cite{Lec1}. It is not 
clear whether or not the presence of life is incompatible with a large radio 
emission of the planet.
\end{enumerate}
Most of the
observable quantities just mentioned are modulated by a phase factor 
$\phi (t)$ with a period $P$ due to the orbital revolution of the planet. 
They are related to the intrinsic charateristics of 
the planetary system by the relations summarized in Table 1.\\

\begin{center}
{\bf Table 1.} Relations between the observable quantities \\and instrisic
characteristics of the planetary system.
\end{center}
\begin{center}
\begin{tabular}{llllll}
$\Delta \theta $& = &$\left( \frac{a}{D}\right) \times \phi (t)$ & = & 0.2 $arc$ $sec$ & \\
 & & & &  & \\
$\Delta \alpha $& = & $\left( \frac{M}{M_{\star}}\right) \left(\frac{a}{D}\right) \times \phi (t)$ & = & 0.6 $\mu as$ & \\
 & & & &   & \\
$\Delta m$ & = & $A\left( \frac{R}{a}\right) ^2 \times \phi (t)$ & = & 25 $mag$ & \\
 & & & &   & \\
$\Delta R_V$ & = & $\left(\frac{M{\rm sin}i}{M_{\star}}\right) \left( {\frac{GM_{\star}}{a}}\right) ^{1/2}  \times \phi (t) $ & = & 0.1 $m$ $s^{-1}$ & \\
 & & & &   & \\
$\Delta T$ & = & $\left(\frac{a}{c}\right) \left(\frac{M}{M_{\star}}\right)$& = &  3 $msec$ & \\
 & & & &   & \\
$\eta $ & = &$\left( \frac{R}{R_{\star}}\right) ^2$ & = & $10^{-4}$ $mag$ & \\
 & & & &   & \\
$D_{occ}$ & = & $\left( \frac{P}{\pi}\right) \left( \frac{R_{\star}}{a}\right)$ & 
= &  15 $hrs$ & \\
 & & & &   & \\
$DT$ & = & $\left( \frac{a}{c}\right) \times \phi (t)$ & = &  8 $min$  & \\
 & & & &   & \\
$A_G$ & $\leq$  & 4$\left( \frac{\sqrt{GMD/c^2}}{R_{\star}}\right)$
& $\leq $ &  1 $mag$ & ($D$ = 1 $kpc$) \\
 & & & &   & \\
$T_G$ & $\approx$ & 4$\sqrt{GMD/c^2}/V$  & $\approx $ &  4 $hrs$  & 
($V$ = 100 km s$^{-1}$)\\
\end{tabular}
\end{center}
\mbox{}\\
The numbers in the right column are for an Earth-sized planet with an albedo 1
at $a=1$ $AU$ from its parent star with 1 solar mass at 5 $pc$ from the Sun
when the phase factor $\phi(t)$ is equal to 1.
For other planetary systems, these numbers scale in an obvious way with
the planet and star parameters. They must be compared
with the present or forthcoming performances of the instrumentation.
We can use this comparison
by classifying the methods of detection according to whether they are adapted
to inner planets, outer planets or all planets. The result is shown in the
Tables 2 and 3 . The third column gives the physical parameters which can be derived
from the observed quantites. The fourth column mentions the astrophysical 
artefacts
which, independently of any systematic {\it instrumental} error, can mimic,
at least before more extensive investigations, the presence of a planet.

The most common methods, namely imaging, astrometry and accelerometry, listed
in Table 2, are in fact sensitive 
only to the presence of outer or giant 
planets (as far as current searches and sufficiently
close to completion projects are concerned). There
are naturally ambitious projects to detect Earth-like inner planets with 
these methods
(see for instance \cite{Ang1}) but 
they have
no chance to become operational before at least a couple of decades.\\

\begin{center}
{\bf Table 2.} Detection methods suited only for outer or giant planets.
\end{center}
\begin{center}
\begin{tabular}{|l|l|l|l|}\hline
{\bf Method} & {\bf Observables} & {\bf Accessible} & {\bf Physical} \\
 & & {\bf parameters} & {\bf artefacts} \\ \hline \hline
imaging &  $\Delta \theta$, $\Delta m$, $P$ & $AR^2$, $a$,  $i$, &  \\ \hline
astrometry  & $\Delta \alpha$, $P$, $i$ & $M$, $a$, $i$ &  \\ 
(optical)  &  &  &  \\ \hline
astrometry  &  $\Delta \alpha$, $P$, $i$ & $M$, $a$, $i$ &  activity \\
(radio)  &  &  & (radio) \\ \hline
accelerometry   & $\Delta V_R$, $P$ & $M{\rm sin}i$, $a$ &  chromospheric \\
 &  &  &  activity ? \\ \hline
\end{tabular}
\end{center}
\mbox{}\\

As for the methods valid for all planets, the most efficient  is the timing 
method. Unfortunately it is
only valid for pulsars and is thus helpless for main sequence stars. The
gravitational amplification method has the disadvantage that it can reveal
the presence of a planet only once (during the single transit due to the proper 
motion of its parent star) and does not allow by itself for subsequent
observations of a detected planet; 
but in very extensive surveys this method could
provide statistical informations on the number of planetary systems. 
\newpage
\begin{center}
{\bf Table 3.} Detection methods suited for all planets.
\end{center}
\begin{center}
\begin{tabular}{|l|l|l|l|}\hline
{\bf Method} & {\bf Observables} & {\bf Accessible} & {\bf Physical} \\
 & & {\bf parameters} & {\bf artefacts} \\ \hline \hline
radio& $F_R$, $P$ ? & $F_R$, $a$ ? &  other \\
emission  &  &  &  sources ? \\ \hline
IR  excess & $F_{IR}$, $\lambda _{max}$ & $R$, $a$ & circumstel. \\ 
 & & & ring \\ \hline
occultations  & $\eta$, $D_{occ}$, $P$ & $R$, $a$, $C$\cite{Sch3}, $i$  & \\ \hline
gravitat.  & $A_{G}$, $\Delta T$ &  $M$, $a_{min}$ ? & \\
amplif.  &  &  &  \\ \hline
timing &  $\Delta t$, $P$ & $M{\rm sin}i$, $a$ &  \\ \hline
\end{tabular}
\end{center}
\mbox{}\\

The 
occultation method is finally the only one which can detect inner planets
around main sequence stars in the near future and can give some of their 
characteristics ($R$, $a$, $i$); it could even subsequently be used to study
the planet atmospheric composition C \cite{Sch3}.

\section{What is effectively done around the world.}
There are many projects and programs for the detection of extrasolar planets. 
Most of them are under development. Some of them have already started. 
Since we are concerned only in the detection of habitable planets, exluding 
therefore the outer or giant planets, I will restrict myself in the current efforts
in the later case. They are listed in Table 4.

\subsection{Ideas, projects and developments.}

\subsubsection{Occultations}
It is worth to give some more details about the occultation method since it
is the only one by which we can hope
to detect habitable planets and measure some of their characteristics in a 
near future. There are two type of
projects. In the space projects \cite{Borucki,STARS},
the aim is to monitor photometrically thousands of G-stars to search for 
inner planets. From the ground, because the photometric precision is then only 
$\approx 10^{-3}$,  the detection of terrestrial planets
is possible only for dwarf (dM) 
stars \cite{Sch2}, for which the transit depth is $10^{-3}$. 
One can increase the efficiency of a random search by choosing stars whose
axis has been predetermined by the comparison of its rotation period, given by
$V_{rot}R_{\star}$, and its projected rotation velocity $V_{rot}$sin$i$ 
\cite{Doy}.

The method finds some improvements when the
target star is an eclipsing binary. There are then several favourable 
characteristics:

- the main advantage is that the chances for correct orientation of the
planet orbit are close to 100\% \cite{Sch1}. This advantage is strenghtened
by the  precession, with a period $P_{pr}=16\cdot P/3(a/a_{sep})^2$, 
of the planet 
orbit around the binary \cite{Sch4}\footnote{$a_{sep}$ being the separation 
of the binary}: it forces
an occultation to occur in any case at least once per half precession period.

- furthermore the binary nature of the central star gives a very characteristic
shape to the occultation curve: there is not only a double occulation
\cite{Sch1}, but it consists of a longer and a shorter occultation since the
motion of the planet is in the same direction as the star for one of the 
components and
in the opposite direction for the other one \cite{Sch1}; 
in addition the ephemerides
of the second occultation can  be predicted with precision from the first one
since the orbital motion of the binary is completely known. This
makes a double occultation difficult to be confused with any 
artefact.

\subsubsection{Other methods}
The Table 4 does not include the timing
searches because there is no timing program specifically aimed for planet 
detection. Almost every timing campaign on pulsars has the potential to detect
planets.

On the other hand I should also mention the spectroscopic search,
although no group has specific plans to search for terrestrial planets with 
this method. Indeed, while a terrestrial planet gives only a reflex velocity
of 0.1 m/s for a solar-type star, this velocity is 0.5 m/s for an
habitable planet orbiting a dM star; 
with a precision of 1 m/s as it is envisaged today on a $m_V=10$ star 
\cite{Con},
it is not hopeless to detect Earth-like planets for these stars with a 10-m
class telescope. Let me remark that for dM stars terrestrial planets have 
orbital periods of the order of weeks and not of years as for Jupiters around
suns; this should render the detection easier. As for the habitability of
planets around dM stars, it has first been argued that the tidal locking of
the planet (which is very close to the star) forbidds the presence of liquid 
water. But 3-D 
atmospheric circulation may prevent the water evaporation.
In addition the internal friction due to
tidal forces can be an extra source of heating.

The lensing method could marginally detect Earth-like planets, but it has two
disadvantages: 1/ it detects the presence of a given planet only once, as a 
single event who will never happen again; 2/ it offers no possibility to 
measure the individual characteristics of the planet, such as mass, inclination
of the orbit or distance 
to the star; it only gives a statistical distribution of a combination of the 
mass and the distance to the star projected onto the sky.
\begin{center}
{\bf Table 4.} Future projects.\\
\begin{tabular}{|l|l|l|l|}\hline
{\bf Method} & {\bf Group} & {\bf Descript.} & {\bf Status}\\ \hline \hline
imaging & Angel\cite{Ang1} & ground adapt. opt. & design only\\
 & & $>$ 6 meters & \\
 & & & \\
 & Angel \cite{Ang2} & space 16 m & design \\
 & & interferom. & \\
 & & & \\\hline
accelerometry & Connes & laser calibr., $\sigma V_R=1$ m/s & tests\\ 
 & & & \\\hline
occultations & FRESIP \cite{Bor} & dedicated  satellite. &  NASA proposal \\
 & & & \\
 & STARS & satellite  dedicated &   ESA proposal\\
 & & to stellar sismology& \\
 & & & \\ \hline
radio & Paris- & direct radio emissions &  tests \\ 
emission & Kharkov \cite{Zarka}& & \\ \hline\hline
\end{tabular}
\end{center}
\mbox{}\\

\subsection{Current observations.}

Since a few years an effective searches of extrasolar
planets with With the occultation method has started in the frame of the `TEP' 
(Transists of Extrasolar Planets) network of mutlisite observations \cite{Deeg}.
This network currently concentrates on eclipsing pairs of 
dwarf cool stars (dM stars), for which the depth of the transit due to 
a terrestrial planet is sufficient to be detectable from the ground
\cite{Schneider5}.

\subsection{What has been found}
It is worth mentioning here all the extra-solar planets that have been
detected as of January 1996.\\

\begin{center}
{\bf Table 6.} Detected extra-solar planets\\

\begin{tabular}{|l|l|l|l|l|}\hline
{\bf Star} & {\bf M}$\times${\bf sin}{\it i}  & {\bf distance to 
star} (AU) & {\bf Method} & {\bf Ref.}\\ \hline \hline
PSR 1257+12 & 0.15 (M$_{\oplus}$ & 0.19 & Timing & \cite{Wol} \\
            & 3.4  & 0.36 & & \\
            & 2.8  & 0.47 & &xxx\\ \hline
51 Pegasus & 0.47 M$_{Jup}$  & 0.05 & Acceler. & \cite{Mayor} \\ \hline
47 Ursa Major & 2.8 M$_{Jup}$ & $\simeq 2$ & Acceler. & \cite{Marcy1} \\ \hline
70 Virginis & 6.4 M$_{Jup}$ & 0.5 &  Acceler. & \cite{Marcy2} \\ \hline \hline
\end{tabular}
\end{center}

In the above list, no one is a good candidate for being an habitable planet.
If the temperature of PSR1257 were $10^6$ K, the planet temperature could be 
around 300 K, but a pulsar is probably an hostile environment for life.
The distance of the companion of 70 Vir to its parent star is such that its
temperature is 80$^o$ C, compatible withe existence of liquid water. But the
object is more likely to be a tiny star formed by condensation of a gas cloud,
and thus to be completely gaseous: the reason for that is the rather high 
eccentricity of its orbit. A planet formed by accretion of small bodies
would have a circular orbit since the trajectories of planetesimals in a 
protoplanetary disk are circularised by random collisions. There is nevertheless
one chance left for the existence of liquid water in this system: the
Jupiter-like companion may well have an Earth-like satellite, just as our 
Jupiter has solid satellites such as Io or Ganymede. It therefore deserves
the most attention of astronomers in a near future.

\subsubsection{Search for $O_2$ and $O_3$}
The final goal of detecting habitable planets is to detect signatures of life.
One very promizing way is the detection of molecular abundances, compatible
only with very out of equilibrium chemical reactions, similar to the high
abundance of $O_2$ in the Earth atmosphere due to the chlorophyllian 
photosynthesis \cite{Lov}. It has been suggested to search, by analogy, for
$O_2$ in the optical A band at \mbox{760 nm \cite{Owen}}
 or for $O_3$ at 9.6 $\mu$m
\cite{Burk}. 

The $O_2$ band in the visible can be sarched for  either by
spectro-imaging of the planet or by pure spectroscopy during a planetary
transit \cite{Sch3}. In the first case a spatial mission would be necessary; in 
the second case is is worth to investigate whether the observations
could be made from the ground with a sufficiently large optical reflector
(as no high anguler resolution is necessary for spectroscopy).

In the case of the $O_3$ band at 9.6 $\mu$m,  a satellised plateform is 
required, to prevent absorption by the Earth's atmosphere \cite{Ang1}. A more careful
investigation, like for instance the DARWIN project submitted to the
European Space Agency \cite{Leg},  shows that it is even necessary to go
at 3.5 AU or more to suppress the IR background of the zodiacal light.

\section{Conclusion}

The detection of  habitable planets around other main sequence stars is still a
great astrophysical and instrumental challenge.  
In the present decade there is an increasing amount of
efforts to detect them by several methods. While the search had up to now
been unsuccessful because of instrumental limitations, the first 
discoveries of Jupiter-like planets are an encouragement to planets hunters
and makes the search
for terrestrial planets more pertinent than ever.

\mbox{}\vspace{0.5cm}\\

Updates of this paper are accessible on the World Wide Web at the URL:

 http://www.obspm.fr/planets/encycl.html

\end{document}